\documentclass[aps,prb,reprint,showpacs]{revtex4-1}

\usepackage{t1enc}
\usepackage[utf8]{inputenc}
\usepackage{amsmath}
\usepackage{graphicx}

\begin{document}

\title{Transition from
coherent mesoscopic single particle transport to proximity Josephson current}

\author{A. Geresdi}
\author{A. Halbritter}
\author{G. Mihály}

\affiliation{{Department of Physics, Budapest University of
Technology and Economics and \\
Condensed Matter Research Group of the Hungarian Academy of Sciences, 1111
Budapest, Budafoki út 8., Hungary}}
\date{\today}
\begin{abstract}
Creating variable size nanojunctions between a thin metallic film and a
superconducting tip we study how multiple phase-coherent scatterings enhance the
superconducting correlations at the normal side. By increasing the coherent
volume of carriers the transmission through the interface is smoothly enhanced as
reflected in the zero bias conductance. As the phase-coherent volume reaches
the opposite surface of the thin film a resonator is formed, the conductance
of the interface is dramatically enhanced, and finally a proximity
induced Josephson supercurrent is established. 
\end{abstract}

\maketitle

Recently, mesoscopic superconductivity and the study of hybrid superconducting
nanostructures have attracted special attention. Hybrid structures, composed from
constituents of fundamentally different electronic structure, have found several
applications, including the nanoscale measurement of spin-polarization by Andreev
spectroscopy \cite{R.J.SoulenJr.10021998}, the creation of a Cooper pair splitter
for the study of entangled electron pairs \cite{Cadden-Zimansky2009,
Hofstetter2009}, or the design of superconducting qubits \cite{Nakamura1999,
Makhlin1999, Mallet2009}.

\begin{figure}
\includegraphics[width=\columnwidth]{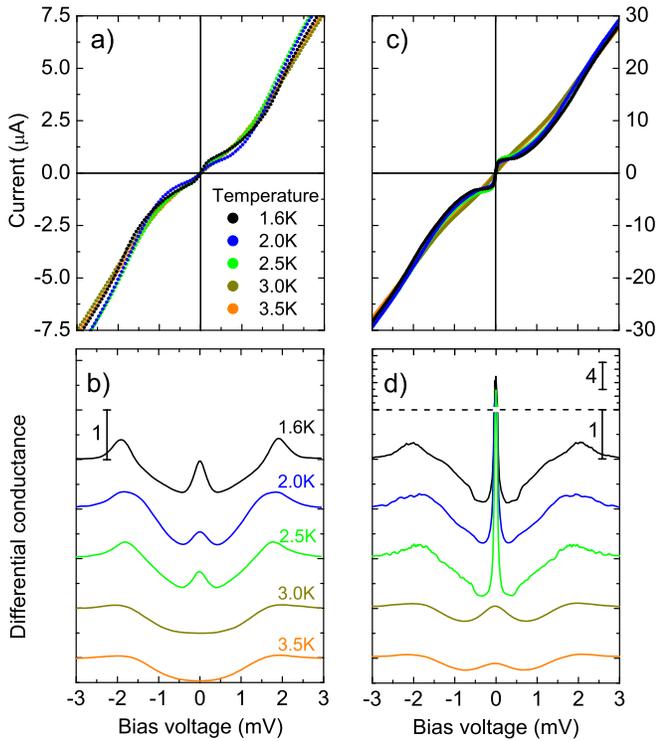}
\caption{\textit{(Color online) Current-voltage characteristics and normalized
differential conductance curves (\(G_\textrm{normal}^{-1}\cdot dI/dV\)) on
Nb-(In,Be)Sb junctions in two different regimes.(a),(b): On smaller junctions
with normal state resistance of \(\approx 400\,\Omega\) the zero bias peak shows
reflectionless tunneling phenomenon. (c), (d): For contacts with larger area
(\(R_\textrm{ normal}\approx 100\,\Omega\)) a huge central peak resembles
Josephson critical current. Note, that the differential conductance curves are
shifted vertically for clarity, and in panel (d) the scale of the region above
the dashed line is squeezed.}}
\label{Fig1}
\end{figure}

The simplest hybrid superconducting nanostructure, a small junction between a
superconductor (S) and a normal metal (N) already features several interesting
phenomena. Andreev reflection, the process of charge transfer at the interface
introduces strongly nonlinear current-voltage characteristics, successfully
described by the theory of Blonder, Tinkham and Klapwijk in ballistic NS
junctions \cite{PhysRevB.25.4515}. For low transparency tunnel junctions the zero
bias conductance -- being determined by the transfer of double electron charges
by Andreev reflection -- is negligible, the current does not increase until the
bias voltage reaches the superconducting gap. Diffusive point contacts exhibit
more complex behaviour. Even for very low junction transparency, a finite
conductance near zero bias can appear due to the particles being reflected at the
NS interface and scattered back from the normal metallic side. The coherent
superposition of trajectories bouncing between the NS junction and the diffusive
medium enhances the transparency of the contact \cite{PhysRevLett.69.510}. The
limit is the full transparency for pair current, i. e. the zero bias conductance
can be as large as twice the normal state conductance:
\(G(V=0)/G_\textrm{normal}=2\) (Ref.~\onlinecite{PhysRevB.64.224513}). At finite
bias voltage this coherent superposition is destroyed, thus a narrow zero bias
peak is observed in the differential conductance, known as reflectionless tunneling
phenomenon.

Reflectionless tunneling has been widely studied on
metal-insulator-superconductor point contacts prepared by lithography. The
materials used on the normal metal side include thin evaporated metallic films
\cite{PhysRevLett.71.1907} or degenerately doped semiconductors
\cite{PhysRevLett.67.3026, PhysRevB.50.4594, PhysRevB.62.9831} and even
nanopatterned structures \cite{PhysRevLett.100.147202}. Similar results were
acquired on more complex layer systems \cite{PhysRevB.65.100508}. The wide range
of the systems proves that reflectionless tunneling is a general phenomenon in NS
junctions, which emerges if the contact radius is larger than the mean free path
of the electrons, but smaller than the phase diffusion length, \(l_m < r <
L_\phi\). 

In order to get a better insight of the various characteristic length scales, we
study both the contact size and the temperature dependence of the zero bias
nonlinearity in diffusive NS junctions using a mechanically controlled point
contact technique. We demonstrate a marked crossover between the coherent
mesoscopic single electron transport and the formation of quasi-bulk proximity
superconductivity. This transition is reflected by the appearance of a clear
Josephson effect. Such critical current in NS systems was predicted in earlier
calculations based on the time-dependent Ginzburg-Landau
framework \cite{Geshkenbein, PhysRevB.35.4669, *PhysRevB.42.8682}, and observed
in the experiments of N.~Agrait, J.~G.~Rodrigo, and S.~Vieria
(Ref.~\onlinecite{PhysRevB.46.5814}).

In this Report we demonstrate that the proximity induced supercurrent can be
enhanced by a special resonator geometry. Our detailed study of the temperature
and contact-size dependence shows, that Cooper pair formation occurs if both the
diameter of the contact and the phase diffusion length defined by the temperature
become larger than the thickness of the normal layer.

As a model material we investigated (In,Be)Sb that was used as nonmagnetic
reference sample in our previous study on spin-polarized transport of (In,Mn)Sb
magnetic semiconductor. These systems are well characterized both by bulk
measurements and by Andreev spectra \cite{Wojtowicz2004325, Csontos2005,
PhysRevLett.100.107201, PhysRevB.77.233304}. Typical thickness of the MBE grown
samples was \(230\,\)nm. Hole concentration, as calculated from earlier Hall
measurements \cite{Wojtowicz2004325} is \(n=1.4 \times
10^{20}\,\textrm{cm}^{-3}\), while the low temperature resistivity is \(\rho =
0.24\,\textrm{m}\Omega \textrm{cm}\). We note that below \(T \approx 10\,\)K the
resistance of the sample is dominated by the residual resistance, and the
saturation value of the mean-free path is \(l_m=9\,\)nm.

Point contacts between the sample and the mechanically sharpened Nb tip are
formed by means of screw thread mechanism and piezo actuation. This tuning
mechanism allowed us to form stable point contacts in a controlled manner with
variable contact size. The contact barrier strength can be acquired by fitting
the \(I-V\) characteristics of ballistic NS junctions using the BTK theory. For
(In,Be)Sb we have obtained \(Z=3 \pm 0.5\) which gives a barrier transparency of
\(\Gamma_\textrm{NS}=\left( 1+Z^2\right)^{-1} \approx 0.1\).  The contact
diameter is deduced from the contact resistance by the Wexler equation
\cite{0370-1328-89-4-316, PhysRevB.70.054416} using the bulk material
parameters \cite{vurgaftman:5815, Wojtowicz2004325} and the barrier transparency.

\begin{figure}
\includegraphics[width=\columnwidth]{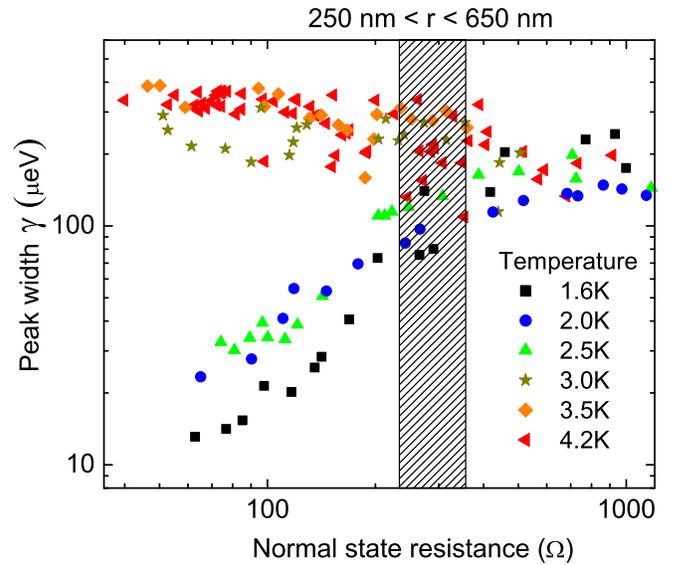}
\caption{\textit{(Color online) Width of the zero bias peak as a function of
contact resistance at different temperatures. Dashed area shows the
transition region in the contact resistance and the corresponding contact diameter.}}
\label{Fig2}
\end{figure}

By fine tuning the piezo displacement stable metallic contacts can be produced
with various sizes ranging from the single atom scale to contact diameters with a
few microns. Here we discuss mostly the diffusive (\(r>l_m\)) point contacts,
measurements in the ballistic limit are analyzed elsewhere
\cite{PhysRevB.77.233304}. The bias dependence of the differential conductance
\(\textrm{d}I/\textrm{d}V\) was investigated for several contacts of different
radius. Typical sets of differential conductance curves and \(I-V\)
characteristics are shown in Fig.~\ref{Fig1}. The curves corresponding to a
smaller junction [Fig.~\ref{Fig1}(a)-(b)] demonstrate a smooth growth of the zero
bias peak resembling reflectionless tunneling phenomenon as the temperature is
decreased. In contrast, for larger contact area [Fig.~\ref{Fig1} (c) and (d)] we
find a sharp increase of the peak amplitude instead of the smooth variation seen
before. At the lowest temperatures the \(\textrm{d}I/\textrm{d}V\) exhibits a
huge central peak which is far beyond what is expected from the phenomenon of
phase coherent multiple scattering; it rather resembles to Josephson critical
current characteristic of superconductor--insulator--superconductor (SIS')
junctions \cite{Josephson1962251, Baselmans1999}.

In order to get more insight into the above features we have measured a large
number of contacts varying the contact size and temperature. The width of the
measured zero-bias peak, \(\gamma\) is plotted against the contact normal state
resistance in Fig.~\ref{Fig2}. For smaller contacts with normal state resistance
larger than \(300\,\Omega\), no significant temperature dependence of \(\gamma\)
is observed. For larger junctions, however, a clear transition is seen at
\(T^\star\approx 2.5 - 3.0\,\)K: at higher temperature \(\gamma\) takes a constant value
of \(\approx 300\, \mu\)V, while at lower temperature it decreases with no apparent
lower limit. The contact resistance at which this transition is observed (dashed
region in Fig.~\ref{Fig2}) can be converted to contact size, revealing a critical contact radius of
\(r^\star\approx 250 - 650 \,\)nm. It is worth noting here that the main source
of error in the size determination is the uncertainty of the contact barrier
strength, \(Z\).

\begin{figure}[!h]
\includegraphics[width=\columnwidth]{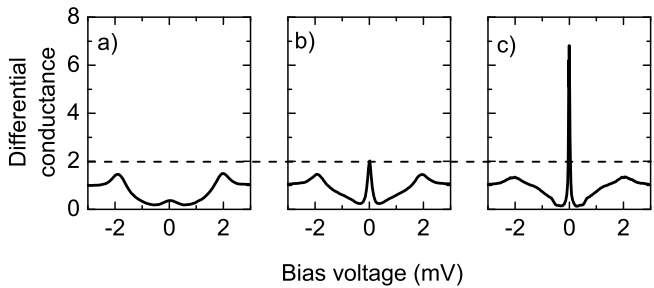}
\caption{\textit{Evolution of the zero bias peak with increasing contact size at
\(T=1.6\,K\). Dashed lines indicate the upper theoretical limit for the peak
amplitude in NS junctions, $G_\textrm{max}(V=0)=2\cdot G_\textrm{normal}$. The
differential conductance curves are normalized to the normal state resistances of
the contacts, which are: (a) \(994\,\Omega\), (b) \(250\,\Omega\), (c)
\(130\,\Omega\), respectively.}}
\label{Fig3}
\end{figure}

\begin{figure}
\includegraphics[width=\columnwidth]{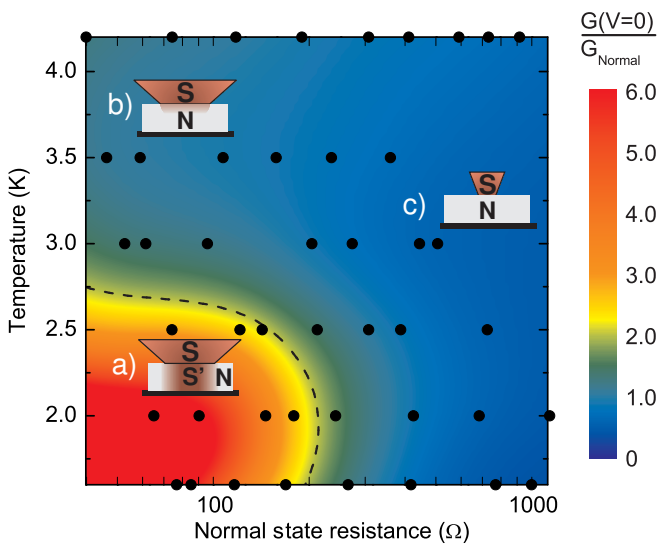}
\caption{\textit{(Color online) Two dimensional plot of the ratio of the zero
bias conductance and the normal state conductance as the function of the point
contact resistance and the temperature. Measured data points are indicated by
black dots. The threshold value of \(G(V=0)/G_\textrm{normal}=2\) is indicated by
dashed line. The three characteristic regions denoted by (a), (b) and (c) are
discussed in the text.}}
\label{Fig4}
\end{figure}

In Fig.~\ref{Fig3} we demonstrate the evolution of the zero bias peak as the function of
the contact size at the lowest measured temperature. The amplitude of the peak is
not only increasing with increasing contact radius, but finally it highly exceeds
the largest possible value for NS junctions, \(G(V=0)_\textrm{max}=2G_N\). This
effect is studied for various contact resistances and temperatures in Fig.~\ref{Fig4} by plotting
the ratio of the zero bias and normal state conductance, and indicating the
threshold value of \(G(V=0)/G_N=2\) with dashed line. Similarly to Fig.~\ref{Fig2} a clear
transition is observed: at low enough temperatures and large enough contact sizes
the amplitude of the zero bias peak exceeds the limit for NS junctions. The critical
temperature and contact radius coincide with those in Fig.~\ref{Fig2}.

Discussing the experimental observations, first we emphasize that both the width
and the amplitude of the zero bias peak indicate a rather sharp transition at a
critical temperature and critical contact radius. For small contacts and high
temperatures zero bias peaks with finite width and moderate amplitudes are
observed, which are attributed to reflectionless tunneling phenomenon. However,
for \(T<T^\star\) and \(r>r^\star\)  the width of the peak abruptly decreases and its height
grows highly above the theoretical upper limit for NS junctions. These features
are characteristic of the formation of Josephson-current, i.~e.~supercurrent
flowing through a barrier between two superconducting regions. The appearance of
a peak narrower than \(k_B T\) is also a clear indication for the formation of
proximity superconductivity in the normal side.

Next we analyze the geometrical constrictions present in the system. The size of
the region where coherence is preserved is described by the phase coherence
length,
\begin{equation}
L_\phi = \sqrt{\frac{\hbar D}{\Delta E}},
\label{Lphi}
\end{equation}
where \(D=0.019\,\textrm{m}^2/\textrm{s}\) is the diffusion constant and \(\Delta
E\) is
the possible energy difference between electron-hole pairs. At zero bias the
dephasing is determined by the temperature \(\Delta E \sim k_B T\). Substituting
the critical temperature of \(T^*=2.5\,\)K a phase coherence length of
\(L_\phi(T^\star)=250\,\)nm is obtained at the crossover temperature, which
coincides with the thickness of the sample. This implies that proximity
superconductivity builds up as the phase coherent scattering region reaches the
bottom surface of the sample, and the density of the phase coherent electron-hole
pairs is enhanced by surface backscattering. The critical contact radius,
\(r^\star\approx 250 - 650 \,\)nm shows good agreement with the thickness
of the sample, indicating that the critical coherent volume for the Cooper pair
formation is reached when the radius of the contact becomes larger then the thickness of
the normal region.

Based on these considerations the parameter space of our measurements can be
divided into three characteristic regions. If the contact size is larger than the
sample thickness, then most of the trajectories can bounce back and forth between
the NS junction and the bottom of the sample. If this entire region is phase
coherent, i.~e.~the trajectories do not loose phase information during several
backscattering events, a proximity superconductivity builds up, and thus
Josephson effect is observed [region (a) in Fig.~\ref{Fig4}]. In this region
calculations based on a step-like function of the superconducting order parameter
are not to be used anymore, and self-consistent methods \cite{PhysRevB.62.10226,
PhysRevB.64.014512} are necessary to obtain proper transport properties. In
contrast, only a small number of trajectories are scattered back coherently if
the phase coherent region does not reach the bottom of the sample because the
\emph{average} pathway yields to a phase loss. In this case the conventional
reflectionless tunneling is observed even for very large junctions [region (b)].
If the contact radius is smaller than the thickness of the sample, the normal
region can be treated as an infinitely thick electrode, thus no crossover
temperature is observed and a small correction is found [region (c)] similarly
to region b). In these two latter cases, the width of the zero bias peak is mostly determined
by the temperature, i.~e.~at \(eV>k_B T\) phase coherence is destroyed by the bias
voltage.

We note that in the proximity superconducting region (a) the low transparency NS
interface is replaced by a Josephson junction, however a new mesoscopic normal
-superconductor interface builds up between the proximity superconducting region,
and the normal electrode. Here the transparency of the interface is large, but
due to the small size of this interface and the 2D character of the sample a
finite contact-resistance is still present. Based on simple estimations the
spreading resistance of the thin film sample is expected to be \(\approx
10\,\Omega\), which is indeed observed in our measurements as a bottom limit of
zero bias resistance. It is important to note that our findings were
reproducible for different lateral positions on the sample. \(I-V\) curves that
were taken during pushing and retracting the tip were found to be identical.

The scheme of the proximity induced Josephson effect enhanced by the geometrical
constrictions present in the system is in good agreement with our experimental
findings. However, additional experiments can further verify the coherent nature
of the huge zero bias peak, like detection of Shapiro steps under microwave
irradiation or observation of Fraunhofer pattern in applied magnetic field. A
systematic study of the layer thickness could also give insight into the details
of the resonance condition.

In conclusion, we have studied the transport through variable size nanoscale NS
junctions in a point contact geometry, by touching a thin metallic film with a
superconducting tip. By increasing the contact size we have induced a clear
transition from reflectionless tunneling phenomenon due to single particle
mesoscopic interference effects to a Josephson supercurrent due to the
condensation of Cooper pairs at the normal side. This transition is generated by
forming a phase coherent resonator region between the junction and the opposite
interface of the metallic film as both the diameter of the junction and the phase
coherence length are enhanced above the thickness of the normal layer.

\begin{acknowledgments}
The authors are grateful to J.K.~Furdyna and T.~Wojtowicz for supplying the
(In,Be)Sb sample studied in this work. This work was supported by the  New
Hungary Development Plan under project ID: TÁMOP-4.2.1/B-09/1/KMR-2010-0002 and
by the Hungarian Research Funds OTKA under grants No. 72916 and No. 76010.
A.~H.~is a grantee of the Bolyai János scholarship.
\end{acknowledgments}


%

\end{document}